\documentclass[aps,twocolumn,showpacs,preprintnumbers,amsmath,amssymb,nofootinbib,superscriptaddress,showkeys,prl]{revtex4}

\usepackage{epsfig}
\usepackage{graphicx}

\begin{document}

\title{Low energy universality and scaling of Van der Waals forces}
\author{A. Calle Cord\'on} 
\email{alvarocalle@ugr.es}
\author{E. Ruiz
  Arriola}
\email{earriola@ugr.es} 
\affiliation{Departamento de
  F\'{\i}sica At\'omica, Molecular y Nuclear, \\ Universidad de Granada,
  E-18071 Granada, Spain.}
  \date{\today}

\begin{abstract} 
\rule{0ex}{3ex} At long distances interactions between neutral ground
state atoms can be described by the Van der Waals potential $ V(r) =
-\sum_{n=6}^\infty C_n /r^n $.  In the ultra-cold regime atom-atom
scattering is dominated by s-waves phase shifts given by an effective
range expansion $ p \cot \delta_0 (p) = -1/\alpha_0 + r_0 p^2/2 +
\dots $ in terms of the scattering length $\alpha_0$ and the effective
range $r_0$.  We show that while for these potentials the scattering
length cannot be predicted, the effective range is given by the
universal low energy theorem $ r_0 = A + B/\alpha_0+C/\alpha_0^2 $
where $A$,$B$ and $C$ depend on the dispersion coefficients $C_n$ and
the reduced di-atom mass.  We confront this formula to about a hundred
determinations of $r_0$ and $\alpha_0$ and show why the result is
dominated by the leading dispersion coefficient $C_6$.  Universality
and scaling extends much beyond naive dimensional analysis estimates.

\end{abstract}

\pacs{34.10.+x; 34.50.Cx; 33.15.-e; 03.75.Nt} 
\keywords{Atom-atom scattering, Van der Waals forces,
Scattering length, Effective range, Renormalization}

\maketitle




Van der Waals (VdW) forces appear ubiquitously in many contexts of
atomic, molecular, nuclear and particle physics. They account for long
range dipole fluctuations between charge neutral atomic and molecular
systems~~\cite{Boschke:1980} with implications on the production of
Bose-Einstein condensates of ultra-cold atoms and
molecules~\cite{2004cucq.book.....W}.  The intermediate range
nucleon-nucleon interaction due to two pion exchange also exhibits
this VdW behaviour based on chiral symmetry~\cite{Tarrach:1977xw}
providing a justification for the liquid drop model of
nuclei~\cite{PavonValderrama:2005wv}. The short distance gluon
exchange interaction between (colour neutral) hadrons also display
this kind of interaction~\cite{Feinberg:1979yw,Fujii:1999xn}.  Van der
Waals forces, however, diverge when naively extrapolated to short
distance scales~\cite{Case:1950an,Frank:1971xx}. The
study of such problems in a variety of situations will certainly shed light on
the usefulness of renormalization ideas within the specific context of
quantum mechanics~(see e.g. Ref.~\cite{PavonValderrama:2007nu}).

Fundamental work for neutral atoms was initiated in
Refs.~\cite{PhysRevA.48.546,Gao:1998zza,PhysRevA.59.1998} (see also
\cite{2009PhRvA..80a2702G}),within a quantum-defect theoretical
viewpoint. In this letter we systematically show that these simplified
approaches work and analyze why they succeed. VdW forces are extremely
simple in this case and are described
by the potential
\begin{eqnarray}
V (r) = - \sum_{n=6}^\infty \frac{C_n}{r^n} \, ,
\label{eq:Cn}
\end{eqnarray}
where $C_n$ are the VdW coefficients which are computed {\it ab
  initio} from intensive electronic orbital atomic structure
calculations~(see e.g. Ref.~\cite{KMS-data} for a
compilation). Usually, only the terms with $n=6,8,10$ are retained
although the series is expected to diverge asymptotically, $C_n \sim
n!$~\cite{1952PhRv...86...92B}. The impressive calculation in Hydrogen
up to $C_{32}$~\cite{2005PhRvA..71c2709M} exhibits the behaviour $C_n
\sim (1/2)^n n!$ at relatively low $n$-values. The
potential~(\ref{eq:Cn}) holds for distances much larger than the
ionization length $l_I = \hbar /\sqrt{ 2 m_e I }$ ($I$ is the
ionization potential) which usually is a few a.u.
\begin{table}
\caption{\label{tab:table_g} Van der Waals length $R_6 =( 2 \mu C_6
  /\hbar^2)^\frac14 $, and the coefficients $g_1$ and $g_2$ defined by
  the reduced dimensionless potential $2 \mu V(r) R_6^2 /\hbar^2
  \equiv - x^{-6} \left[ 1 + g_1 x^{-2}+ g_2 x^{-4} + \dots \right] $
  with $x=r/R_6$. We use results from
  Refs.~\cite{2001PhRvA..63e2704D,2003JChPh.119..844P,
    1998PhRvA..58.4259M, 2006JETP..102..195P, 2005PhRvL..94r3201W,
    1996PhRvA..54.2824Y} }
\begin{ruledtabular}
\begin{tabular}{|c|c|c|c|}
{\rm Atoms} &$R_6(a.u.)$&$g_1 (10^{-2})$&$g_2 (10^{-4})$ \\ \hline 
Li-Li     &  64.9214    &    1.42458    &    2.97874    \\ \hline 
Na-Na     &  89.8620    &    0.92320    &    1.11369    \\ \hline 
K-K       & 128.9846    &    0.64780    &    0.49784    \\ \hline 
Rb-Rb     & 164.1528    &    0.45647    &    0.23370    \\ \hline 
Cs-Cs     & 201.8432    &    0.36544    &    0.13983    \\ \hline 
Fr-Fr     & 215.0006    &    0.27362    &    0.09526    \\ \hline 
Li-Na     &  73.2251    &    1.25605    &    2.17183    \\ \hline 
Li-K      &  84.2285    &    1.18374    &    1.79689    \\ \hline 
Li-Rb     &  88.0587    &    1.18572    &    1.70555    \\ \hline 
Li-Cs     &  92.8950    &    1.21364    &    1.68241    \\ \hline 
Na-K      & 106.5708    &    0.80600    &    0.80155    \\ \hline 
Na-Rb     & 115.3377    &    0.74528    &    0.65923    \\ \hline 
Na-Cs     & 123.2277    &    0.73874    &    0.61148    \\ \hline 
K-Rb      & 142.8292    &    0.56543    &    0.37106    \\ \hline 
K-Cs      & 154.2909    &    0.53903    &    0.32152    \\ \hline 
Rb-Cs     & 180.8480    &    0.41520    &    0.18654    \\ \hline 
Be-Be     &  43.3013    &    2.54953    &    6.70303    \\ \hline
Mg-Mg     & 72.3589     &    1.26414    &    1.60399    \\ \hline
Ca-Ca     & 111.4907    &    0.84584    &    0.65057    \\ \hline
Sr-Sr     & 148.9023    &    0.55117    &    0.27632    \\ \hline
Ba-Ba     & 189.4340    &    0.41692    &    0.15260    \\ \hline
Cr-Cr     &  91.2731    &    1.22821    &    $-$        \\ \hline \hline 
H-H       &  10.4532    &   17.51760    &   423.45426    \\ \hline 
He-He     &  10.1610    &    9.35937    &  117.94642     
\end{tabular}
\end{ruledtabular}
\end{table}
In the Born-Oppenheimer approximation the quantum mechanical problem
consists of solving the Schr\"odinger equation for the two atoms apart
a distance $r$,
\begin{eqnarray}
-u_k'' + U(r) u_k + \frac{l(l+1)}{r^2}u_k= k^2 u_k \, , 
\label{eq:schr} 
\end{eqnarray} 
where $U(r) = 2 \mu V(r)/\hbar^2 $ is the reduced potential, $\mu =
m_1 m_2 /(m_1 +m_2) $ the reduced di-atom mass, $k=p/\hbar =2\pi/
\lambda $ the wavenumber, and $u_k(r)$ the reduced wave function. For
our purposes, it is convenient to write the reduced potential in VdW
units,
\begin{eqnarray}
U(r) = - \frac{R_6^4}{r^6} \left[ 1 + g_1\frac{R_6^2}{r^2}+ 
g_2\frac{R_6^4}{r^4} + \dots  \right] \, ,  
\label{eq:C6-C8-C10}
\end{eqnarray} 
where $ R_6 = ( 2 \mu C_6 /\hbar^2)^\frac{1}{4}$ is the VdW length
scale and $g_n = (2 \mu C_{6+2n} /\hbar^2) R_6^{-4-2n}$ for $n \ge
1$. In Table~\ref{tab:table_g} we display numerical values for several
di-atomic systems which are extremely small in the VdW units $g_1 \sim
10^{-2}$ and $g_2\sim 10^{-4}$ with the exception of H and He. Thus,
we may anticipate a dominance of the $C_6$ term in the calculations at
low energies. This will be systematically quantified below.

Using the superposition principle for positive
energy scattering s-waves we decompose the general solution as 
\begin{eqnarray}
u_k (r) = u_{k,c} (r)  +  k \cot \delta_0 (k) \,  u_{k,s} (r)   \, ,  
\label{eq:sup_k}
\end{eqnarray} 
with $ u_{k,c} (r) \to \cos (k r) $ and $ u_{k,s} (r) \to \sin (k r)
/k $ for $r \to \infty $ and $\delta_0(k) $  the scattering phase
shift for the $l=0$ angular momentum state.  The potential given by
Eq.~(\ref{eq:Cn}) is both long range and singular at short
distances. At short distances, the De Broglie wavelength is slowly
varying and hence a WKB approximation
holds~\cite{Case:1950an,Frank:1971xx}, yielding for $ r \to 0 $
\begin{eqnarray}
u_k (r) &\to & C \left(\frac{r}{R_n}\right)^{\frac{n}{4}} \sin\left[
\frac{2}{n-2} \left(\frac{R_n}{r}\right)^{\frac{n}2-1} + \varphi_k \right] \, ,  \label{eq:WKB} 
\end{eqnarray} 
where $R_n= (2\mu C_n/\hbar^2)^{1/(n-2)}$ corresponds to the highest VdW scale
considered in Eq.~(\ref{eq:Cn}). The
phase $\varphi_k $ is in principle arbitrary. On the other hand, at
low energies one has the effective range
expansion~\cite{1963JMP.....4...54L}
\begin{eqnarray}
k \cot \delta_0 (k)= - \frac1{\alpha_0} + \frac12 r_0 k^2 + v_2 k^4
\log (k^2) + \dots
\label{eq:ere} 
\end{eqnarray} 
where $\alpha_0$ is the scattering length, and $r_0$ is the effective
range which can be calculated from  
\begin{eqnarray} 
r_0 &=& 2 \int_0^\infty dr \left[ \left(1-\frac{r}{\alpha_0} \right)^2-
u_0 (r)^2 \right] \, . 
\label{eq:r0_singlet}
\end{eqnarray} 
Here, the zero energy solution becomes, from Eq.~(\ref{eq:sup_k}), 
\begin{eqnarray}
u_0 (r) = u_{0,c} (r)  - u_{0,s} (r) / \alpha_0  \, ,   
\label{eq:sup_0}
\end{eqnarray} 
where $ u_{0,c} (r) \to 1 $ and $ u_{0,s} (r) \to r $ for $r \to
\infty $, yielding 
\begin{eqnarray} 
r_0  &=&  A + \frac{B}{\alpha_0}+ \frac{C}{\alpha_0^2}  \, ,    
\label{eq:r0_univ} 
\end{eqnarray} 
with $A$,$B$ and $C$ given by 
\begin{eqnarray}
A &=& 2 \int_0^\infty dr ( 1 - u_{0,c}^2 ) \, , \\    
B &=& -4 \int_0^\infty dr ( r - u_{0,c} u_{0,s} ) \, , \\    
C &=& 2 \int_0^\infty dr ( r^2 - u_{0,s}^2 )    \, .   
\end{eqnarray} 
Then, combining the zero and finite energy
wave functions we get for any $r_c>0$ 
\begin{eqnarray}
u_k'(r) u_0 (r) - u_0'(r) u_k(r) \Big|_{r_c}^\infty = k^2
\int_{r_c}^\infty u_k(r) u_0(r) dr \, ,   
\end{eqnarray} 
where $r_c$ plays the role of a short distance cut-off which is
innocuous provided $l_I \ll r_c \ll R_n$. Using 
Eq.~(\ref{eq:sup_k}), Eq.~(\ref{eq:WKB}) and Eq.~(\ref{eq:sup_0})
we then get for $r_c \to 0$
\begin{eqnarray}
\frac1{R_n} \sin (\varphi_k - \varphi_0) &=& k^2 \int_0^\infty dr
\left[u_{0,c} (r) - \frac{1}{\alpha_0} \, u_{0,s} (r) \right]
\nonumber \\ &\times& \Big[ u_{k,c} (r) + k \cot \delta_0 (k) \,
  u_{k,s} (r) \Big] \, .
\end{eqnarray} 
Orthogonality between $u_k$ and $u_0$ requires $\varphi_k=\varphi_0$
in which case, expanding the integrand we get the structure
\begin{eqnarray}
k \cot \delta_0 (k) = \frac{ \alpha_0 {\cal A} ( k) + {\cal B} (k)}{
\alpha_0 {\cal C} (k) + {\cal D} (k)} \, , 
\label{eq:phase_singlet}
\end{eqnarray} 
where ${\cal A}$, ${\cal B}$, ${\cal C}$ and ${\cal D}$ are suitable
functions of $k$.  The interesting feature of Eqs.~(\ref{eq:r0_univ})
and (\ref{eq:phase_singlet}) is that the dependence on the scattering
length $\alpha_0$ and the potential is explicitly disentangled. This
is a universal form of a low energy theorem, which applies to {\it
  any} potential regular or singular at the origin which falls off
faster than $1/r^5$ at large distances.  We can visualize
Eq.~(\ref{eq:r0_univ}) as a long distance (VdW) correlation between
$r_0$ and $\alpha_0$. If the reduced potential depends on a {\it
  single } scale $R$, i.e. $U(r) = - F( r/R) /R^2 $, one gets {\it
  universal scaling relations}
\begin{eqnarray} 
\frac{r_0}{R} = \bar A + \bar B \frac{R}{\alpha_0}+ \bar C
\frac{R^2}{\alpha_0^2} \, ,
\label{eq:r0_univ_scaled} 
\end{eqnarray}
where $\bar A$,$\bar B$ and $\bar C$ are purely geometric numbers
which depend {\it solely} on the functional form of the potential.

\begin{figure*}[ttt]
\begin{center}
\epsfig{figure=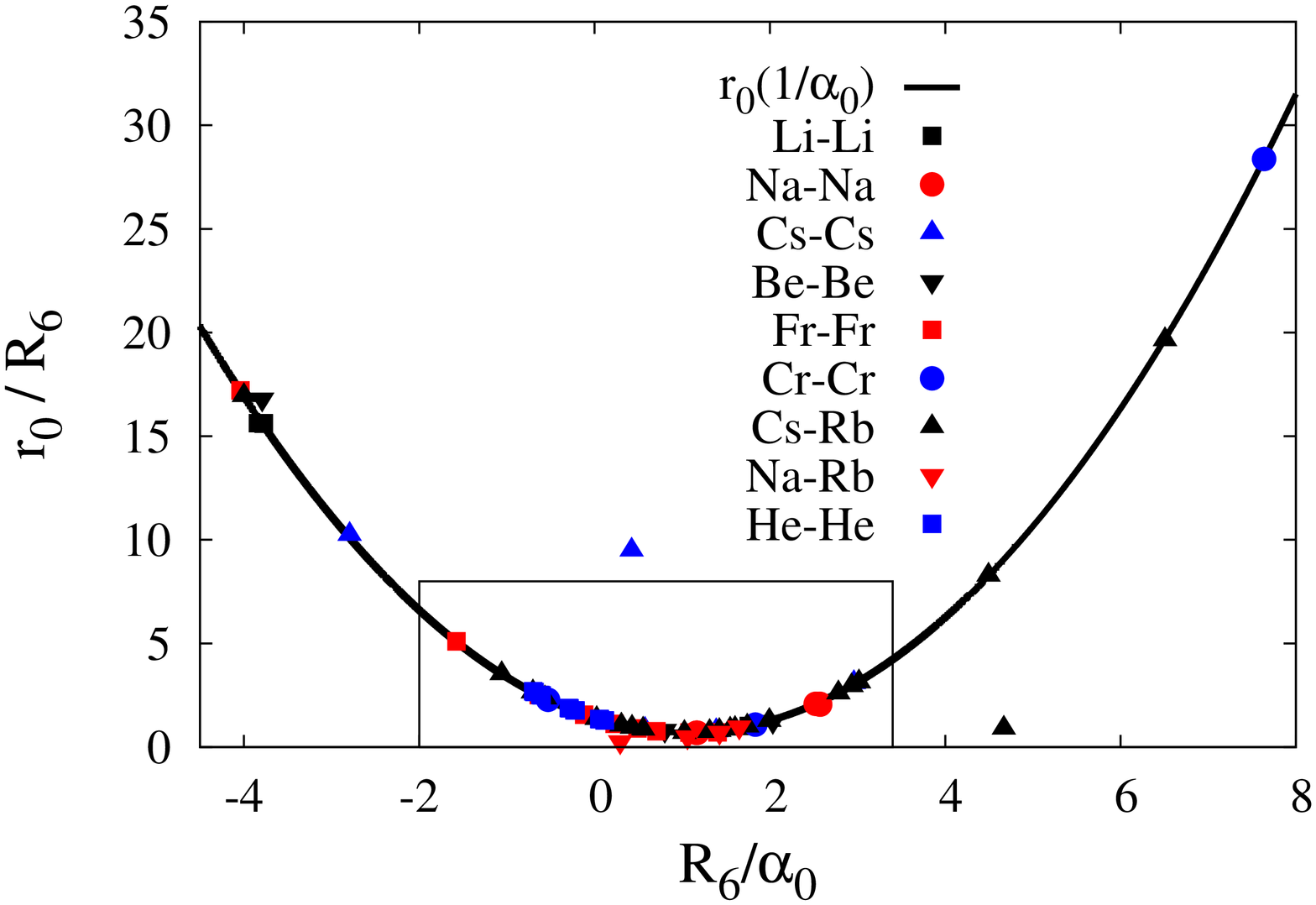,height=5.5cm,width=5cm,angle=270}
\epsfig{figure=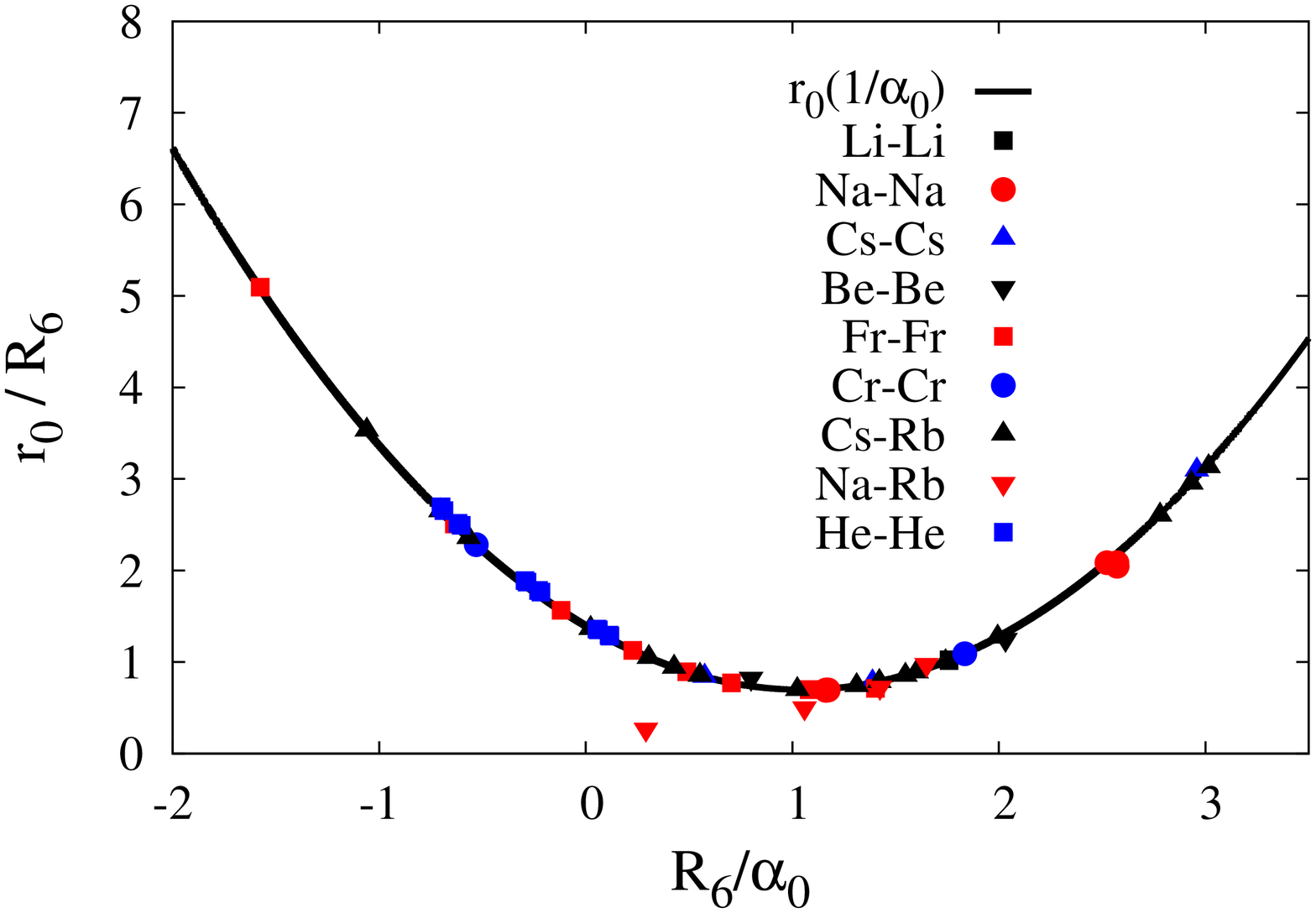,height=5.5cm,width=5cm,angle=270}
\epsfig{figure=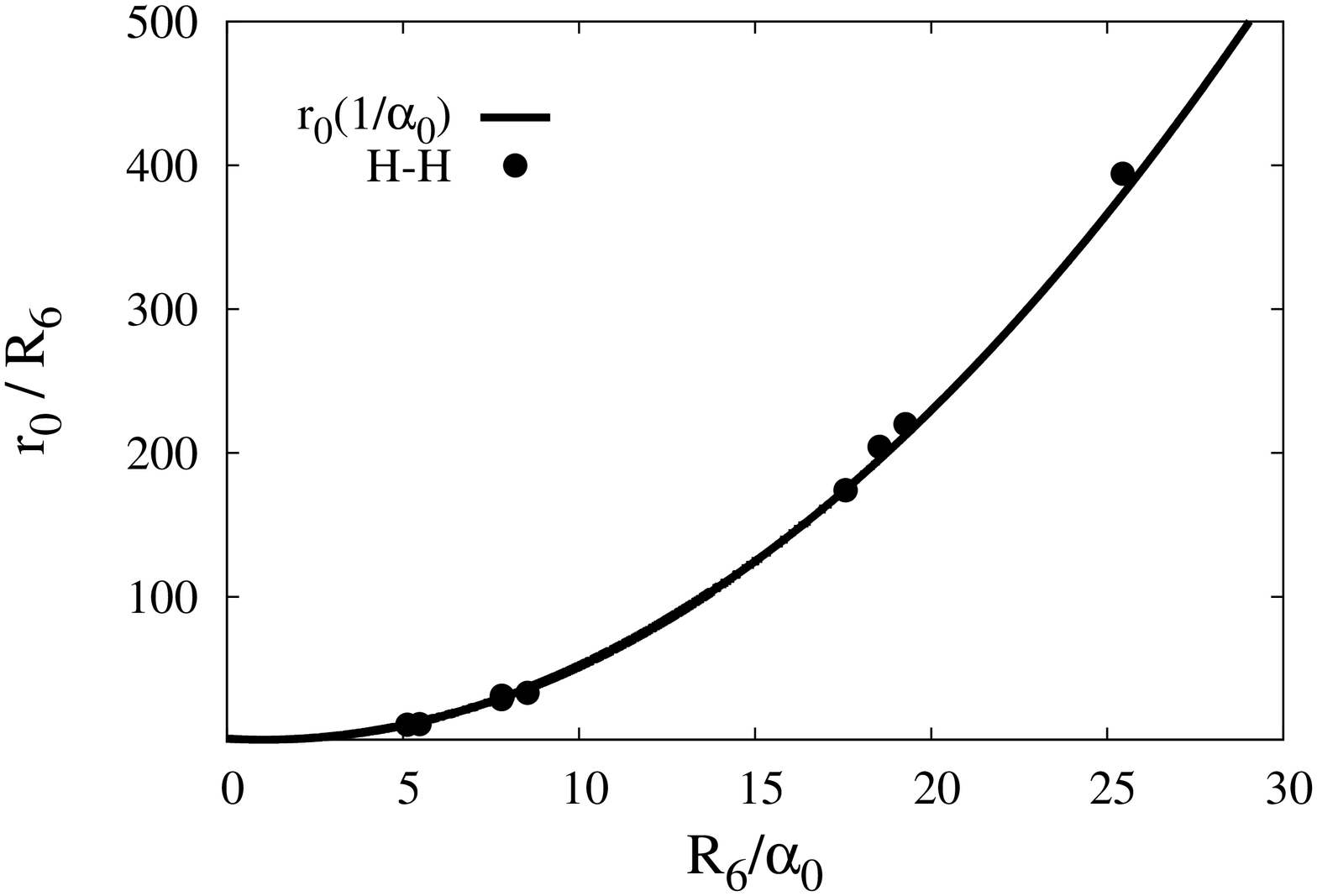,height=5.5cm,width=5cm,angle=270}
\end{center}
\caption{The effective range $r_0$ vs the inverse scattering length
  $1/\alpha_0$ in units of the VdW radius $R_6=(2\mu C_6/\hbar^2)^{\frac14}$
  for different ranges. Points are potential
  calculations~\cite{1996PhRvA..53..234C,1994PhRvA..50..399C,1994PhRvA..50.4827C}
  (Li-Li,Na-Na), \cite{1994PhRvA..50.3177M} (Cs-Cs),
  \cite{2009EPJD...53...27O} (Na-Rb), \cite{2007JPhB...40.3497J}
  (Be-Be), \cite{2003JPhB...36.1085J} (Cs-Rb),
  \cite{2004PhRvA..69c0701P} (Cr-Cr), \cite{2004physics...6027K}
  (Fr-Fr),
  \cite{1986JPSJ...55..801K,1992PhRvA..46.6956J,2006EL.....76..582S,1995PhRvA..51.2626J}
  (H-H), \cite{1995PhRvA..51.2626J} (He-He).  The line corresponds to
  Eq.~(\ref{eq:r0-ggf})~\cite{Gao:1998zza,PhysRevA.59.1998}.}
\label{fig:atoms-curve}
\end{figure*}

For the pure VdW case, $V=-C_6/r^6$, 
the effective range has been computed
analytically~\cite{Gao:1998zza,PhysRevA.59.1998} and in harmony with
the general structure Eq.~(\ref{eq:r0_univ_scaled}) reads
\begin{eqnarray}
\frac{r_0}{R_6} & = & 1.39473 - 1.33333\, \frac{R_6}{\alpha_0} +
0.637318 \frac{R_6^2}{\alpha_0^2} \, ,
\label{eq:r0-ggf}
\end{eqnarray}
In Fig.~\ref{fig:atoms-curve} we confront the prediction for the
effective range to the result of many potential calculations in VdW
units.  As can be vividly seen the agreement is rather impressive
taking into account the simplicity of
Eq.~(\ref{eq:r0-ggf}). Alternatively, and discarding the exceptional
outliers for $\alpha_0$ and $r_0$, we perform a $\chi^2$ fit to the
form Eq.~(\ref{eq:r0_univ_scaled}) for the remaining 82 points and get
$A= 1.31$, $B= -1.57$ and $ C= 0.66$ in good agreement with
Eq.~(\ref{eq:r0-ggf}). To our knowledge, Fig.~\ref{fig:atoms-curve}
represents a completely unforeseen universal correlation supported by
phenomenology.

\begin{figure}[ttt]
\begin{center}
\epsfig{figure=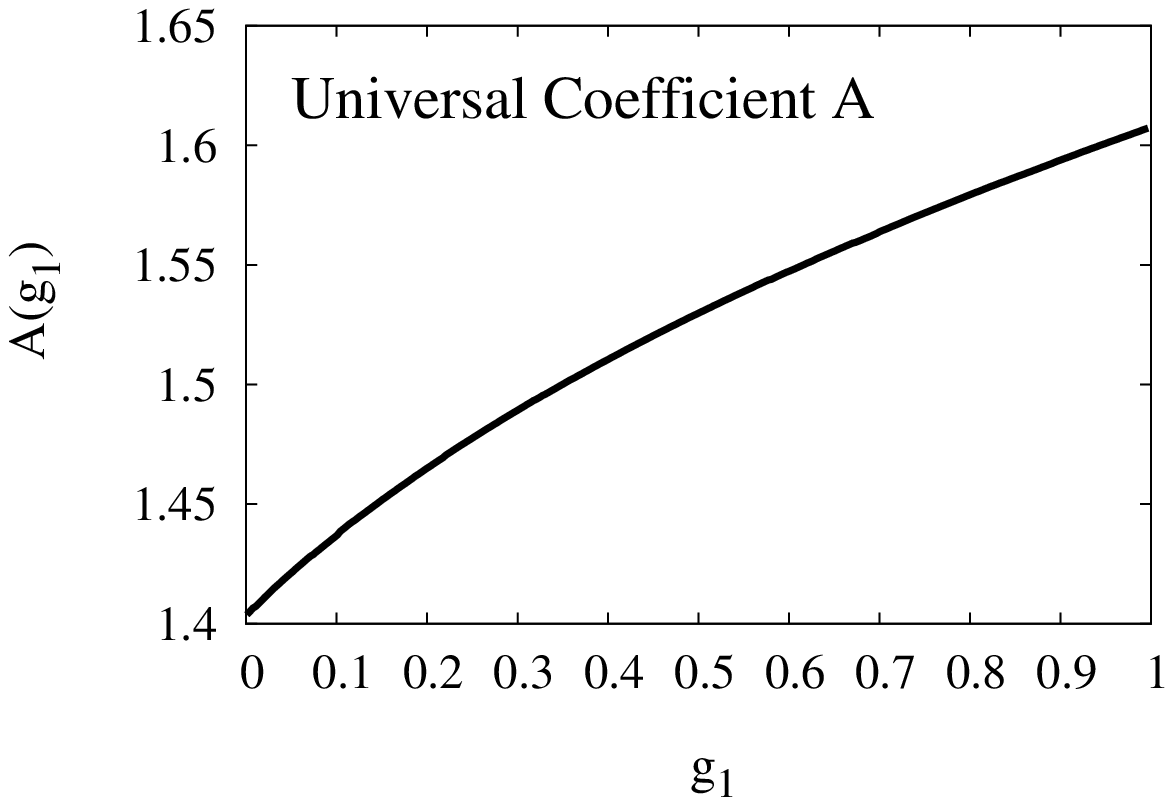,height=3.7cm,width=4cm}
\epsfig{figure=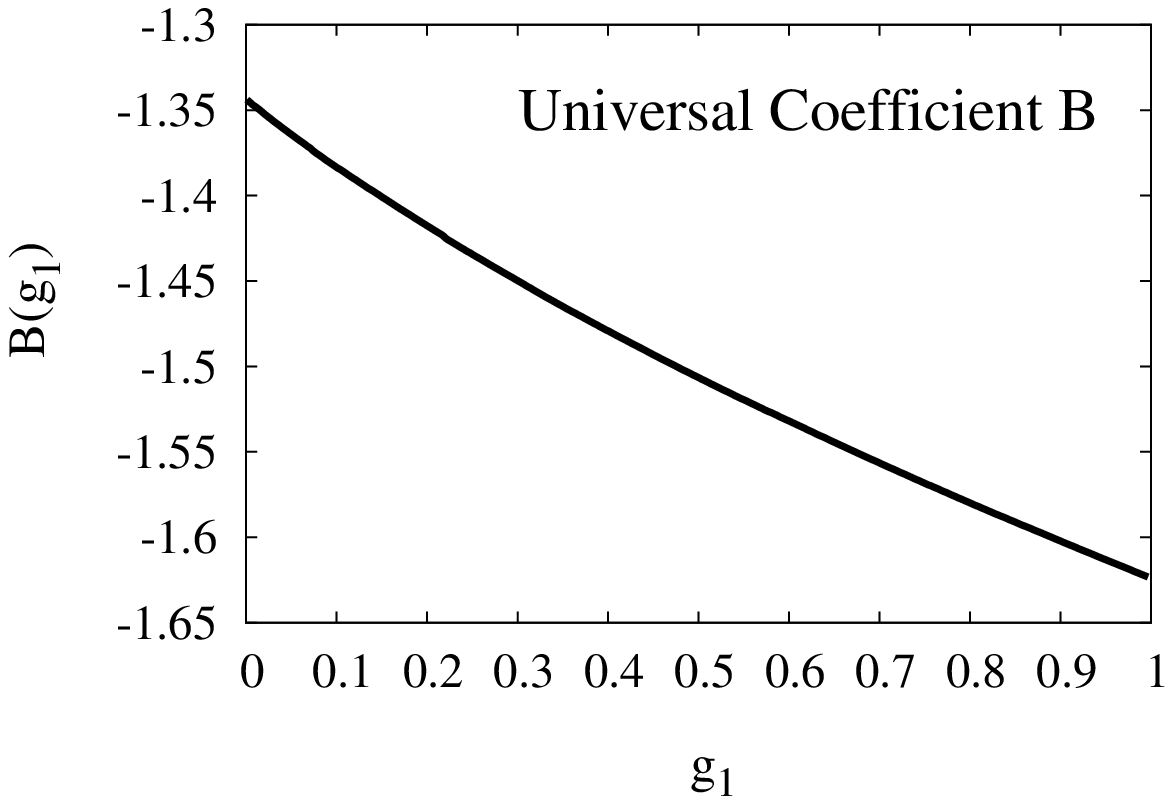,height=3.7cm,width=4cm}
\epsfig{figure=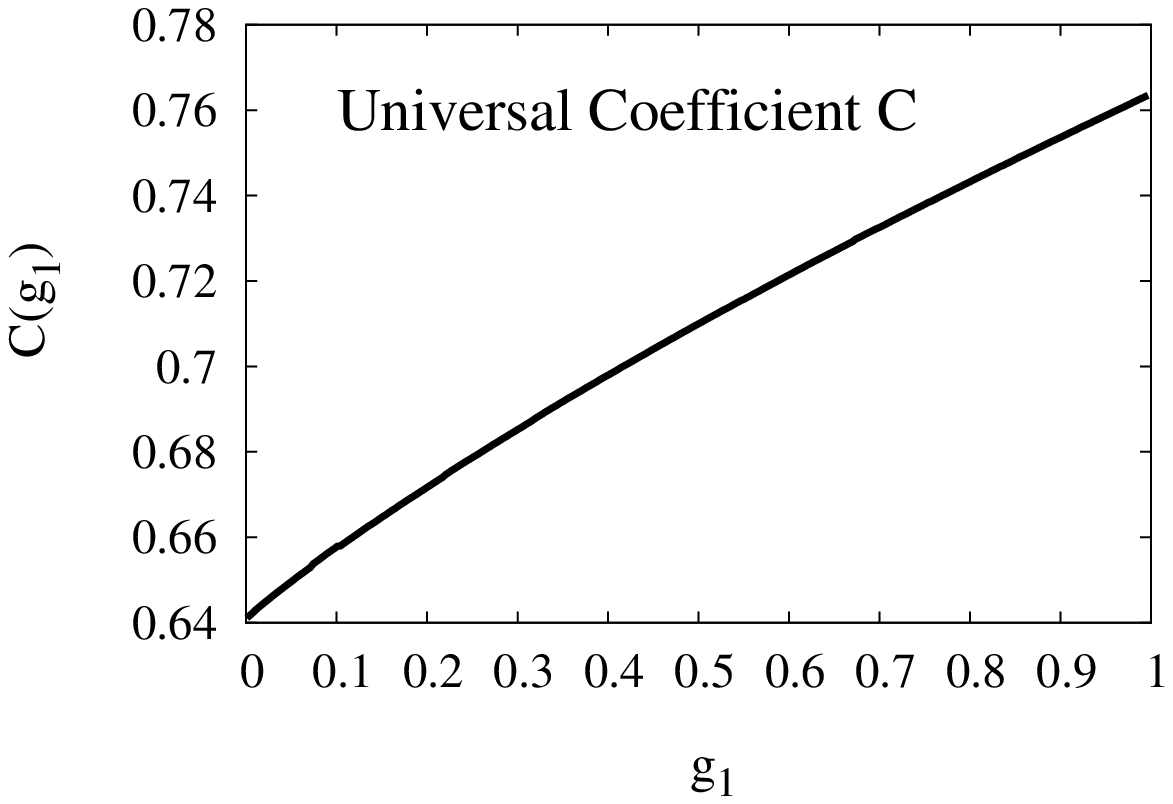,height=3.7cm,width=4cm}
\end{center}
\caption{ The effective range coefficients $A$, $B$ and $C$ (see
  Eq.~(\ref{eq:r0_univ})) as a function of the dimensionless coupling
  $g_1$ representing the $1/r^8 $ correction to the VdW potential
  $1/r^6$.}
\label{fig:power}
\end{figure}
We analyze the robustness of this agreement by showing in
Fig.~\ref{fig:power} the effect on the effective range coefficients,
$A (g_1)$, $B(g_1)$ and $C(g_1)$ due to adding a $1/r^8$ term.  From
Table~\ref{tab:table_g} we see that mostly $g_1 \sim 10^{-2} $, in
which case tiny changes are expected from
Fig.~\ref{fig:power}. Actually, the smallness of the deviations
suggests using perturbation theory.  If we expand the full solutions
of the $1/r^6$ potential at small $k$, $u_k(r) = u_0 (r)+ k^2 u_2(r) +
\dots$, the change in the effective range due to inclusion of a
$\Delta U(r) \sim 1/r^8$ potential keeping $\alpha_0$ {\it fixed }
reads~\cite{PavonValderrama:2005wv,Valderrama:2009ei}
\begin{eqnarray}
\Delta r_0 = 4 \int_{r_c}^\infty \Delta U(r) u_0 (r) u_2(r) \, dr  
\end{eqnarray} 
where $r_c \sim g_1 R_6 \ll R_6$. The leading contribution is
determined by the short distance behaviour of $u_0(r)$, see
Eq.~(\ref{eq:WKB}), and we find $u_2(r) \sim (r/ R_6)^4 u_0(r)$
yielding $ \Delta r_0 \sim g_1 R_6 \log g_1 + {\cal O} (g_1) $. The
logarithmic enhancement in $g_1$ can indeed be observed in
Fig.~\ref{fig:power} by the tiny curvature.

\begin{figure}[ttt]
\begin{center}
\epsfig{figure=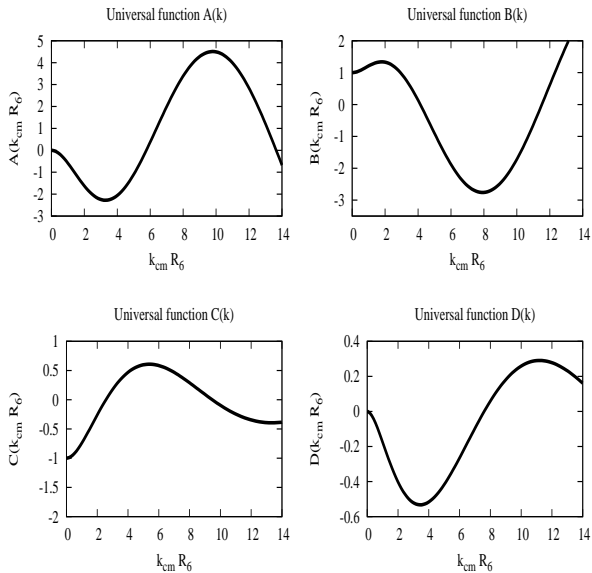,height=8cm,width=8cm,angle=270}
\end{center}
\caption{\small The universal functions in units of 
  $R_6$ defined by $ 2\mu V(r) = -R_6^4/r^6$ which allow to determine
  the phase shift if the scattering length is also known, see 
  Eq.~(\ref{eq:phase_singlet}).}
\label{fig:ABCD}      
\end{figure}

The $C_6$-dominance is compelling and we show in Fig.~\ref{fig:ABCD}
the universal functions ${\cal A}(k)$, ${\cal B}(k)$, ${\cal C}(k)$
and ${\cal D}(k)$ which in conjunction with $\alpha_0$ allow to
determine the phase-shift from Eq.~(\ref{eq:phase_singlet}). They
scale with $R_6$ and are uniquely determined by the power law $-1/r^6$
{\it once and forever}. We have found that these functions show little
dependence on $g_1$ and $g_2$ at momenta as large as $k R_6 \sim 10$,
a rather unexpected result, hinting that the VdW universality and
scaling extends much beyond the naive dimensional analysis estimate $
k R_6 \approx 1$ or the effective range expansion of
Eq.~(\ref{eq:ere}) truncated with the first two terms.  We note in
passing that although such a truncation suggests a higher degree of
universality, the VdW nature of the interaction prevents using
$(\alpha_0,r_0)$ as fully independent variables in view of
Eq.~(\ref{eq:r0-ggf}) and Fig.~\ref{fig:atoms-curve}. It is
interesting to notice that within a Nuclear Physics context
characterized by short ranged Yukawa potentials stemming from meson
exchange~\cite{CalleCordon:2008cz} the general Eq.~(\ref{eq:r0_univ})
has been exploited as a means to check SU(4) Wigner symmetry in the
Nucleon-Nucleon interaction for the $^1S_0$ and $^3S_1$ channels with
a pattern similar to Fig.~\ref{fig:atoms-curve}.

We conclude by underlining that, when suitably displayed, the
analytical approach to Van der Waals forces pursued in previous
works~\cite{Gao:1998zza,PhysRevA.59.1998} acquires a quite universal
character with indisputable phenomenological success; the leading
$1/r^6$ contribution suffices to accurately describe low energy
atom-atom scattering with just two parameters in a wide energy
range. We naturally expect new universality and scaling patterns to
emerge from systems characterized by power law forces but less
understood such as molecular interactions in the ultra-cold
regime. From a broader perspective we stress that the lack of
dependence of potential model calculations on short distance details,
unveiled from our systematic comparison, is a feature traditionally
built-in by the quantum-defect theory. Its natural counterpart of a
smooth and controllable short distance cut-off dependence complies to
the requirement of renormalizability within a pure quantum-mechanical
framework.

We thank M. Pav\'on Valderrama and R. Gonz\'alez F\'erez for discussions
and M. Gacesa, P. Pellegrini and Z. Pavlovi{\'c} for comments. Work
supported by the Ministerio de Ciencia y Tecnolog\'\i a, contract
no. FIS2008-01143/FIS and Junta de Andaluc{\'\i}a grant no.FQM225-05.


\end{document}